\documentclass[a4paper,fleqn]{cas-dc}
\usepackage[authoryear,longnamesfirst]{natbib}

\def\tsc#1{\csdef{#1}{\textsc{\lowercase{#1}}\xspace}}
\tsc{WGM}
\tsc{QE}

\begin{document}
\let\WriteBookmarks\relax
\def\floatpagepagefraction{1}
\def\textpagefraction{.001}

\shorttitle{Stiffening of matter in quark-hadron continuity: a mini-review}    
\shortauthors{Toru Kojo}  
\title [mode = title]{Stiffening of matter in quark-hadron continuity: a mini-review}  
\author{Toru Kojo}[orcid=0000-0003-2815-0564]
\ead{torukojo@post.kek.jp}
\affiliation{organization={Theory Center, IPNS, High Energy Accelerator Research Organization (KEK)},
            addressline={1-1 Oho}, 
            city={Tsukuba, Ibaraki},
            postcode={305-0801}, 
            country={Japan}}
\affiliation{organization={Graduate Institute for Advanced Studies, SOKENDAI},
            addressline={1-1 Oho}, 
            city={Tsukuba, Ibaraki},
            postcode={305-0801}, 
            country={Japan}}
\affiliation{organization={Department of Physics, Tohoku University},
            addressline={Aoba-yama}, 
            city={Sendai, Miyagi},
            postcode={980-8578}, 
            country={Japan}}

\begin{abstract}
Recent observations of neutron stars, combined with causality, thermodynamic stability, and nuclear constraints, indicate a rapid stiffening of QCD matter at densities slightly above nuclear saturation density ($n_0 \simeq 0.16\,{\rm fm}^{-3}$). The evolution of the stiffening is faster than expected from purely nucleonic models with many-body repulsion. Taking into account the quark substructure of baryons, we argue that the saturation of quark states occurs at $\sim$ 2-3$n_0$, driving quark matter formation even before baryonic cores of radius $\sim$0.5 fm spatially overlap. We describe the continuous transitions from hadronic to quark matter within a quarkyonic matter model in which gluons are assumed to remain confining at densities of interest. To obtain analytic insight into the transient regime, we construct an ideal model of quarkyonic matter, the {\it IdylliQ} model, in which one can freely switch from baryonic to quark descriptions and vice versa.
\end{abstract}

\maketitle

\section{Introduction}

\begin{figure}
    \centering
    \includegraphics[width=0.9\linewidth]{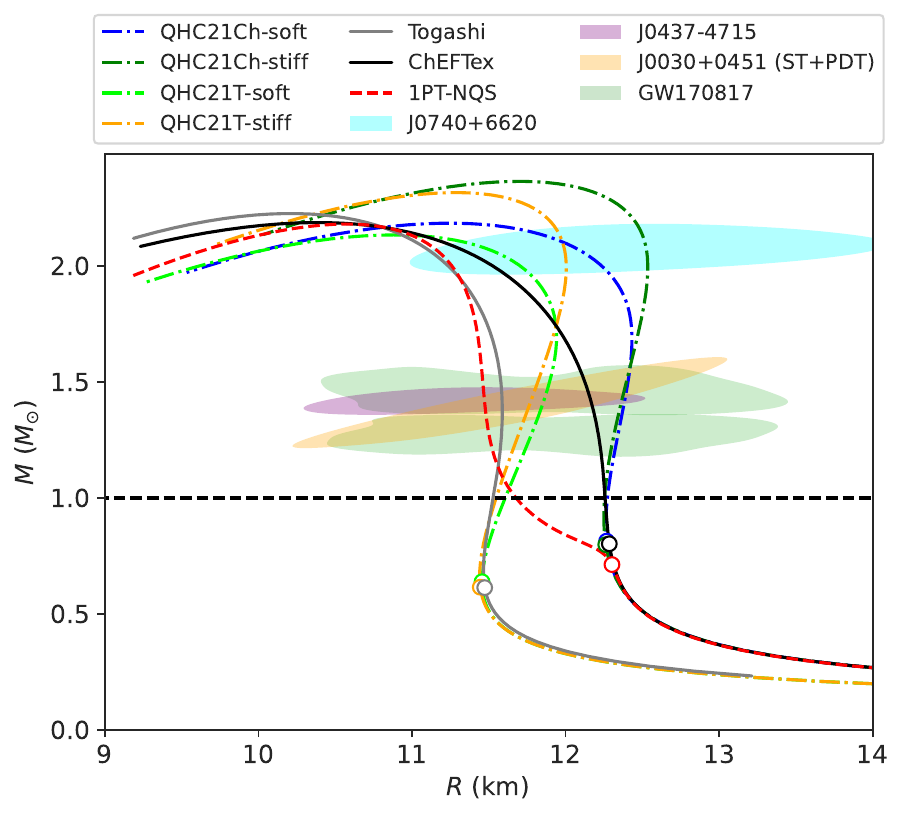}
    \includegraphics[width=0.9\linewidth]{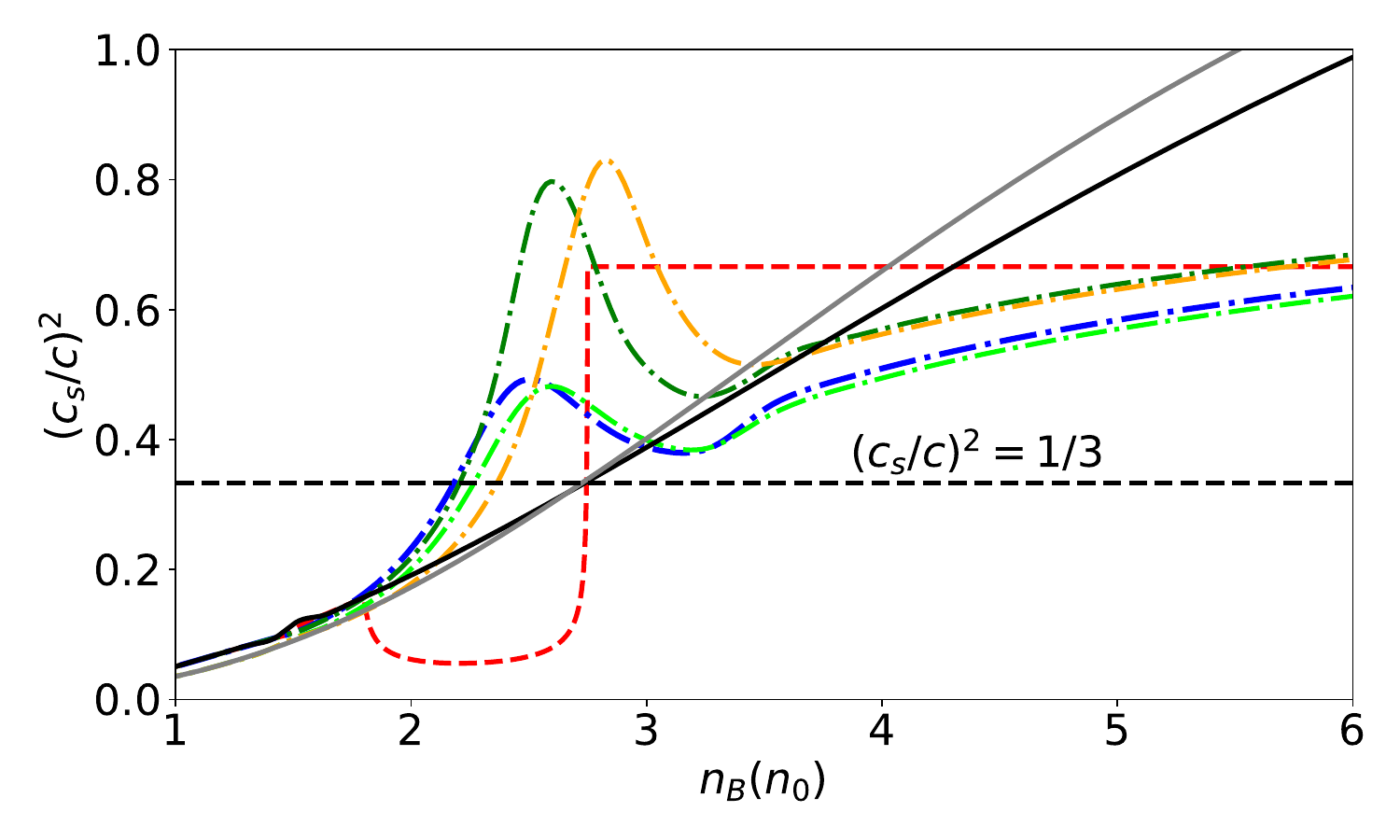}
    \caption{ Examples of $M$-$R$ relations (upper panel) and $c_s^2$ as a function of $n_B/n_0$ (lower panel) for the crossover (QHC21 models \citep{Kojo:2021wax}) and hadronic models (ChEFTex \citep{Hensh:2024onv} and Togashi \citep{Togashi:2017mjp}).
    The bands in the $M$-$R$ plots are from observational constraints with the 68\% confidence interval.
    The hadronic models include three-body forces which stiffen EOS but its extrapolation toward high density violates the causality constraint around $\simeq$ 5-6$n_0$.
    The figures are taken from \cite{Hensh:2024onv}.
    }
    \label{fig:cs2}
\end{figure}

The properties of highly compressed matter are one of the central topics in quantum chromodynamics (QCD).
The most basic quantities are equations of state (EOS) \citep{Oertel:2016bki,Baym:2017whm}. 
They are directly related to the neutron star (NS) structure, especially the mass-radius ($M$-$R$) relations.
The $M$-$R$ relations have one-to-one correspondence with the pressure-energy density ($P$-$\varepsilon$) relations, and can be categorized according to the stiffness of EOS.
For stiff (soft) EOS, $P$ is large (small) at a given $\varepsilon$, and the former (the latter) leads to larger (smaller) radii and masses.

One of the crucial discoveries from NS observations is the rapid evolution of the stiffness.
A useful measure to characterize the density evolution of the stiffening is the sound speed, $c_s = (dP/d\varepsilon)^{1/2}$,
whose behaviors are strongly correlated with the shape of $M$-$R$ curves \citep{Tan:2021nat,Kojo:2020krb}.
Recent NS observations for $1.4M_\odot$ and $2.1M_\odot$ NSs ($M_\odot$: the solar mass), together with causality, thermodynamic stability, 
and nuclear constraints at low density $\sim n_0$ ($n_0 \simeq 0.16\,{\rm fm}^{-3}$: nuclear saturation density) \citep{Drischler:2021kxf}, 
suggest that the stiffness evolves rapidly, as suggested by recent Bayesian inference studies \citep{Han:2022rug,Marczenko:2022jhl,Brandes:2024wpq}.
The rapid stiffening begins to occur slightly above $n_0$, and considerable pressure is developed already at density of 2-3$n_0$.

One of the key questions is how to interpret such rapid stiffening.
In the language of nucleonic matter, stiffening is typically achieved by two- and three-body repulsion, see, e.g., \cite{Akmal:1998cf}. 
We parametrize the EOS as ($a$ is a coefficient universal for kinetic energies)
\begin{eqnarray}
\varepsilon_N (n_B) = m_N n_B + a \frac{n_B^{5/3} }{m_N} + b n_B^N \,,
\label{eq:en_para}
\end{eqnarray}
where the first term is from the mass energy, the second from the non-relativistic kinetic energy, and the third from $N$-body interaction whose coefficient $b$ is taken positive.
This class of matter is very soft unless the interaction dominates over the other terms.
Indeed, computing the pressure $P = n_B^2 \partial (\varepsilon/n_B)/\partial n_B$, we find
\begin{eqnarray}
P_N = \frac{2a}{3} \frac{n_B^{5/3} }{m_N} + b (N-1) n_B^N \,.
\end{eqnarray}
We note that the leading mass term, which dominates the energy density at moderate densities, does not contribute to the pressure. 
The kinetic energy is suppressed by the large nucleon mass. Hence the pressure is dominated by interactions.
If we extrapolate the present parametrization to high density, 
the $N$-body terms become dominant and can even exceed the mass energy. 
In this high density limit, let us keep only the $N$-body term for both the energy and pressure. 
Then we get $P_N(\varepsilon)$ and $c_s^2$ at large $n_B$,
\begin{eqnarray}
P_N \sim (N-1) \varepsilon_N ~~\rightarrow~~~ c_s^2 \sim N-1 \,.
\end{eqnarray}
The squared sound speed $c_s^2$ asymptotically approaches $N-1$; 
in the two-body case, $c_s^2 \rightarrow 1$; in the three-body case, $c_s^2 \rightarrow 2$, 
thereby violating the causality bound $c_s^2 \le c^2$ with $c=1$ being the speed of light in natural units. 
While many-body repulsion offers the stiffness necessary to explain the existence of $2M_\odot$ NS, it presents at least two problems.
First, the importance of two-body and three-body forces raises concerns regarding the convergence of many-body expansion.
Second, the power growth of the stiffness is rather slow and the predicted radius of a $2M_\odot$ tends to be smaller 
than the radius constraint of $2.1M_\odot$ NS \citep{Kojo:2021wax}.

The behavior of nuclear many-body forces is difficult to infer within nuclear models. 
For example in order to avoid the causality violation in models incorporating three-body repulsions, 
we must include more-body forces with negative contributions to moderate the growth of $c_s^2$. 
The cancellations among large positive and large negative terms are theoretically undesirable in constructing theories. 
To address this issue at fundamental level, we need quark descriptions.

\section{Quark descriptions}

The simplest quark EOS is  that of a free, massless quark gas:
\begin{eqnarray}
\varepsilon_q (n) = a' n^{4/3} + \mathcal{B} \,,
\end{eqnarray}
where $a'$ is some constant and $\mathcal{B}$ is the normalization constant for the EOS.
By computing the pressure and eliminating $a'$ in favor of $\varepsilon$, we find $P = \varepsilon/3 - 4\mathcal{B}/3$ from which the conformal limit $c_s^2 =1/3$ follows.
We note that, depending on the value of $\mathcal{B}$, the EOS can be either very stiff or very soft. 
In fact, the maximum mass $M_{\rm max}$ and the corresponding $R|_{ M_{\rm max} }$ can be expressed as functions of $\mathcal{B}$ as
\citep{Witten:1984rs}
\begin{eqnarray}
M_{\rm max}/M_{\odot} &=& 2.03 M_\odot \bigg( \frac{\, 56\,{\rm MeV/ fm^{3}} \,}{ \mathcal{B} } \bigg)^{1/2} \,,
\notag \\
R |_{ M_{\rm max} }&=& 10.7\,{\rm km} \times \frac{M_{\rm max} }{\, 2.0 M_\odot \,} \,.
\end{eqnarray}
With $\mathcal{B} \lesssim 56\, {\rm MeV/fm^3}$, the $2M_\odot$ constraint can be satisfied; relativistic kinetic energy of quarks can produce sufficiently large pressure. 
However, models with larger values of $\mathcal{B}$ are often chosen  to better describe hybrid hadron-to-quark matter transitions.

If we trust that nuclear matter at low density represents the true ground state, the above argument highlights the interplay between nuclear and quark matter.
Instead of considering massless quark matter, we focus here on matter composed of non-relativistic quarks to isolate discussions on the mass reduction at high density.
We assume $M_N \simeq N_c M_q$ where $M_q \simeq 300$ MeV is the constituent quark mass.
The energy density for two-flavor quark matter is parametrized as
\begin{eqnarray}
\varepsilon_q (n_B) = N_c M_q n_B + a \times N_c \frac{\, n_B^{5/3} \,}{M_q} + \cdots \,,
\label{eq:eq_para}
\end{eqnarray}
where the constant $a$ is the same as in Eq.~\eqref{eq:en_para}. 
We take the quark Fermi momentum to be $\sim n_B^{1/3}$ 
since the quark density per color is equal to the baryon density, i.e.,
$n_q^{\rm red} = n_q^{\rm green} = n_q^{\rm blue} = n_B$.
The corresponding pressure is then given by
\begin{eqnarray}
P_q = \frac{2a}{3} \times N_c \frac{n_B^{5/3} }{M_q} + \cdots \,.
\end{eqnarray}
We emphasize that the first term in $\varepsilon_q$ is similar for nucleonic and quark descriptions since $N_c M_q \simeq m_N$.
However,  the coefficients of the kinetic energy terms differ significantly:
they scale as $\sim 1/N_c$ in nuclear models,  while as $\sim N_c$ in quark models.
This results in approximately a factor of $N_c^2 \sim 10$ difference.
Accordingly, the pressure originating from the kinetic energy is roughly ten times larger in quark descriptions than in nuclear ones without many-body repulsion.
To construct a stiff EOS with $P_q \sim \varepsilon_q $, 
a pressure of $O(N_c)$ is required; in nuclear models, this pressure primarily arises from interactions, 
whereas in quark models it can be provided by the kinetic energies. 

Note that, if we take Eqs.~\eqref{eq:en_para} and \eqref{eq:eq_para} literally, we would arrive at one of the following conclusions:
(i) If nuclear forces are neglected in Eq.~\eqref{eq:en_para}, then $\varepsilon_N < \varepsilon_q$ holds because of the large kinetic energy in quark matter descriptions. 
As a result, quark matter never appears, and the EOS remains soft.
(ii) If nuclear repulsive forces are included, then at sufficiently large density $\varepsilon_N$ exceeds $\varepsilon_q$. 
Similarly, the extrapolated $P_N$ eventually becomes greater than $P_q$ at large density.
Before these excesses actually occur, a first order phase transition should take place, 
where the chemical potentials become equal, 
$\mu_B = \partial \varepsilon_N/\partial n_B = \partial \varepsilon_q/\partial n_B$. 
At this point, the matter transforms into quark matter to avoid the increasing energy cost associated with nuclear many-body repulsion.
After the phase transition, the EOS becomes softer,
which may give impressions that quark matter EOS is intrinsically soft.

In the following, we develop arguments showing that nuclear models {\it without} interactions 
are inevitably driven to transform into quark matter as the density 
increases---even before many-body repulsion becomes dominant in the nuclear EOS.
This transition is not merely a consequence of avoiding the energy cost of many-body repulsion; 
rather, 
it is more fundamentally rooted in the fact that baryons are composed of quarks, 
and thus must respect the Pauli principle at the quark level.
During this transformation, changes in the energy density are modest and continuous, while the pressure increases rapidly.
Nuclear many-body interactions are discussed only 
briefly—their primary role is to smooth out the transition by bridging the gap between the nuclear and quark EOS.

\section{A quarkyonic matter model}

Over the past decade, models describing a crossover from hadronic to quark matter have been developed 
as templates that incorporate constraints from astrophysics, nuclear physics, as well as causal and thermodynamic stability conditions 
\citep{Masuda:2012kf,Masuda:2012ed,Kojo:2014rca}.
These models are mostly based on phenomenological interpolation but have successfully extracted some general trends, such as the emergence of a sound speed peak,
the relatively weaker role of strangeness in quark matter compared to hadronic matter, the importance of pairing effects, and others.
Meanwhile, explicit theoretical descriptions of the crossover regime had remained  undeveloped until recently.
Below, we discuss one concrete realization of the crossover based on the concept of quarkyonic matter.

\subsection{Sum rules and quark saturation}

In order to describe a crossover within a unified setup, we begin by considering quarks confined in a single baryon, 
and then extend the discussion to quarks in a multi-baryon system.
Naturally, it is extremely difficult to directly handle quantum many-body states. 
Therefore, we adopt a coarse-grained approach.
We characterize the system by specifying how baryonic and quark states are occupied in the many-body environment.
Denoting the quark momentum distribution within a baryons by $\varphi$, 
and the occupation probabilities of baryons and quarks by $f_B$ and $f_Q$, respectively,
we propose the following simple sum rule (with $\int_p \equiv \int d^3 \bf{p}/(2\pi)^3$) \citep{Kojo:2021ugu}
\begin{eqnarray}
f_Q (q) = \int_{k} f_B ({\bf k}) \varphi ({\bf q} -{\bf k}/N_c) \,,
\end{eqnarray}
where $f_Q$ is the quark distribution function for a given color; explicitly, $f_Q^{\rm red} = f_Q^{\rm green} = f_Q^{\rm blue} \equiv f_Q$.
The RHS is simply represents the sum of quark states originating from baryons.
The average momentum of a quark inside a baryon with total momentum $k$ is given by $k/N_c$, reflecting that the baryon is composed of 
$N_c$ quarks sharing the total momentum.
The normalization of $\varphi$ is $\int_q \varphi (q) = 1$.
We assume the width of the distribution $\varphi $ is $\sim \Lambda \sim 0.2-0.4$ GeV so that the proton radius is $\sim$ 0.5-1.0 fm.

The sum rule has the following general properties:
(i) Integrating over $q$, we find $\int_q f_Q = n_B = \int_k f_B$. 
The total number of quarks per color is equal to the baryon number, as expected.
(ii) If we fix $k$ at a low momentum and take $q\rightarrow \infty$, the asymptotic behavior becomes $f_Q(q) \sim \varphi(q) \int_k f_B(k) = n_B \varphi (q)$.
(iii) In the large $N_c$ limit, assuming $k/N_c \ll \Lambda$, $f_Q$ scales as $f_Q(q) \sim \varphi(q) \int_k f_B(k) = n_B \varphi (q)$.

The last point (iii) warrants special attention. 
The probability $f_Q$ increases with $n_B$, eventually reaching the upper bound of $1$ --- first occurring at $q=0$.
We call it {\it quark saturation} \citep{Kojo:2021ugu}.
At this point, it is no longer valid to treat $k/N_c$ in $\varphi$ as a small parameter.
To clarify this issue, let us consider $q=0$ and perform a change of variables by rescaling $k/N_c = k'$, so that
\begin{eqnarray}
f_Q (0) = N_c^3 \int_{k'} f_B(N_c k') \varphi(k') \,,
\end{eqnarray}
subject to the constraint $n_B = \int_k f_B (k) = N_c^3 \int_{k'} f_B(N_c k')$.
At first glance, the prefactor $N_c^3$ appears to violate the Pauli blocking constraint $f_Q \le 1$,
but it depends on the support of the integrand, which is determined by $f_B$.
If we write the range of $|\bf{k}|$ as $[0, k_{\rm max}]$, 
the corresponding range of $|\bf{k'} |$ shrinks to $[0, k_{\rm max}/N_c]$. 
As a result, the available phase space is reduced by a factor $1/N_c^3$, 
which cancels the prefactor $N_c^3$ in front of the integral.
But once $k_{\rm max}$ becomes $O(N_c)$, the phase space for $k'$ becomes $O(1)$ and the $N_c^3$ prefactor can no longer be cancelled.
To prevent violation of the Pauli principle, we need $f_B (k) \sim 1/N_c^3$ over most of the range $[0, k_{\rm max}]$ \citep{Kojo:2019raj}.
This situation implies that quark states at low momenta become saturated, and the domain of saturation expands with increasing baryon density
As saturation proceeds, baryons are only allowed to occupy low-momentum states with a small probability. 
Meanwhile, at sufficiently high momenta, baryons can become free from quark saturation constraints. 
These considerations lead naturally to the momentum shell picture for the baryonic distribution, 
which we will derive explicitly using an idealized model of quarkyonic matter.

\subsection{IdylliQ model}

We now construct a dynamical model which incorporates the sum rule as a constraint.
To gain analytic insights, we consider the following idealized setup \citep{Fujimoto:2023mzy}:
(i) The model neglects all interactions except those responsible for quark confinement.
Confinement is effectively implemented by requiring that quarks exist only as constituents of baryons.
Under this assumption, the energy density is given by
\begin{eqnarray}
\varepsilon [f_B] = \int_k E_B (k) f_B (k) \,.
\label{eq:en_fB}
\end{eqnarray}
(ii) We ignore the density dependence of the internal quark momentum distribution $\varphi$.
In realistic settings, the width of $\varphi$ in momentum space is expected to shrink as baryons swell in dense matter. 
However, to retain tractability, we fix $\varphi$ throughout.
(iii) We adopt a specific functional form for $\varphi$ that allows for an analytic inversion of the sum rule, enabling us to express 
$f_B$ as a functional of the quark distribution $f_Q$.
Explicitly, we take:
%
\begin{eqnarray}
\varphi (q) = \frac{\, 2 \pi^2 \,}{\, \Lambda^3 \,} \frac{\, e^{-q/\Lambda} \,}{\, q/\Lambda \,} \,,
\end{eqnarray}
which is the inverse of the operator
\begin{eqnarray}
L = - \nabla_q^2 + \frac{1}{\, \Lambda^2 \,} \,,~~~~~ L \big[ \varphi ({\bf q}-{\bf p} ) \big] = (2\pi)^3 \delta (\bf{q} - \bf{p}) \,.
\end{eqnarray}
With this we can express $f_B$ as
\begin{eqnarray}
f_B (N_c q) = \frac{\, \Lambda^2 \,}{\, N_c^3 \,} L \big[ f_Q (q) \big] \,.
\end{eqnarray}
For instance, in a domain where quark states are saturated, i.e., $f_Q (q) = 1$, then
the derivative term vanishes and we obtain $f_B (N_c q) = 1/N_c^3$.

One might question the necessity of assumption (iii), particularly since the chosen Yukawa form of the distribution $\varphi$ may appear unrealistic. 
The rationale, however, lies in our desire to capture the qualitative behavior of $f_B$ at high density.
In this regime, we possess physical intuition grounded in quark-based descriptions—for instance, the emergence of a quark Fermi sea. 
Our interest lies in the densities that are high but still within the vicinity of the crossover regime, 
where such quark-based intuitions are becoming applicable, yet hadronic features remain relevant. 
Conversely, at low density, we have reasonable intuition for $f_B$,
and we can compute $f_Q$ using the sum rule. 
Thus, by adopting assumption (iii), we can bridge the low- and high-density regimes using two complementary descriptions. 
This dual perspective offers a more robust framework than one relying solely on either hadronic or quark degrees of freedom.

\begin{figure}
    \centering
       \vspace{-1cm}
    \includegraphics[width=0.9\linewidth]{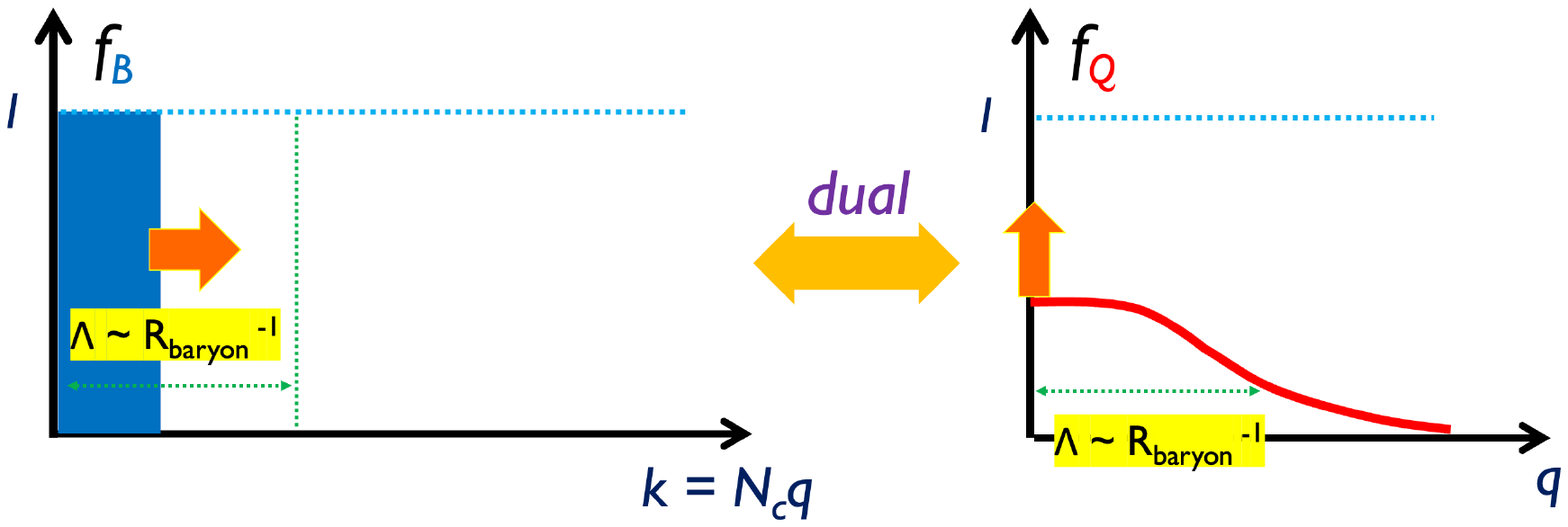}
    \vspace{-1.4cm}
    \caption{ $f_B$ and $f_Q$ before quark saturation.
    }
      \vspace{-1.cm}
    \includegraphics[width=0.9\linewidth]{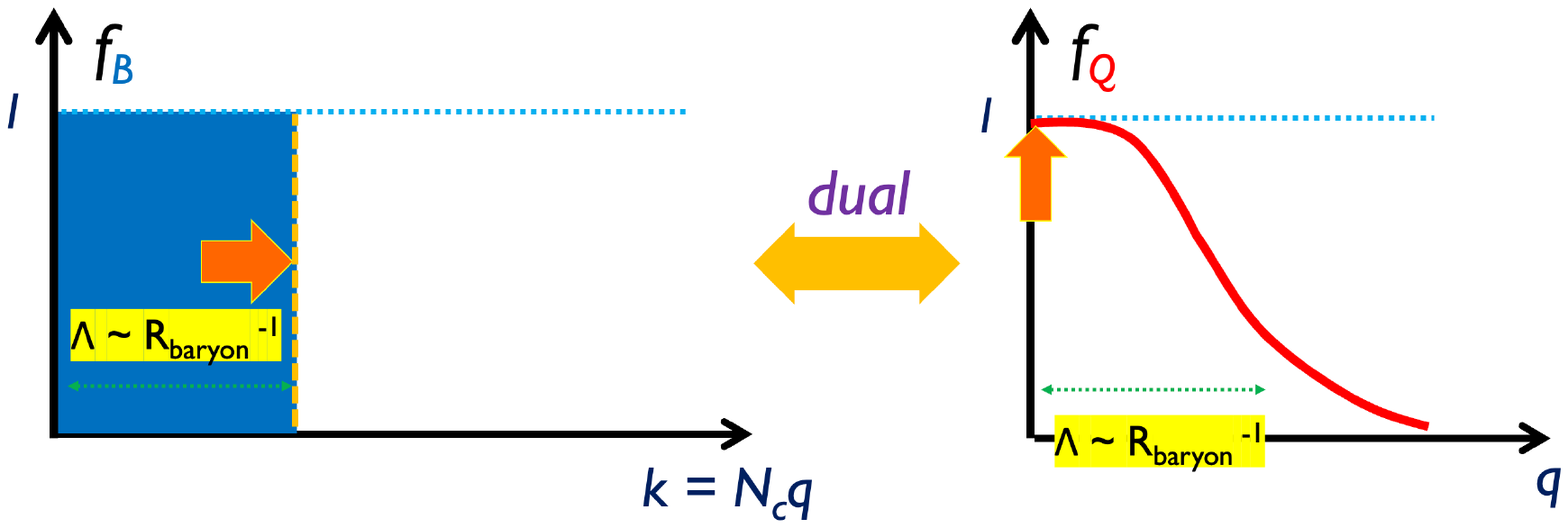}    
        \vspace{-1.4cm}
       \caption{ $f_B$ and $f_Q$ at quark saturation.
    }
            \vspace{-1.cm}
     \includegraphics[width=0.9\linewidth]{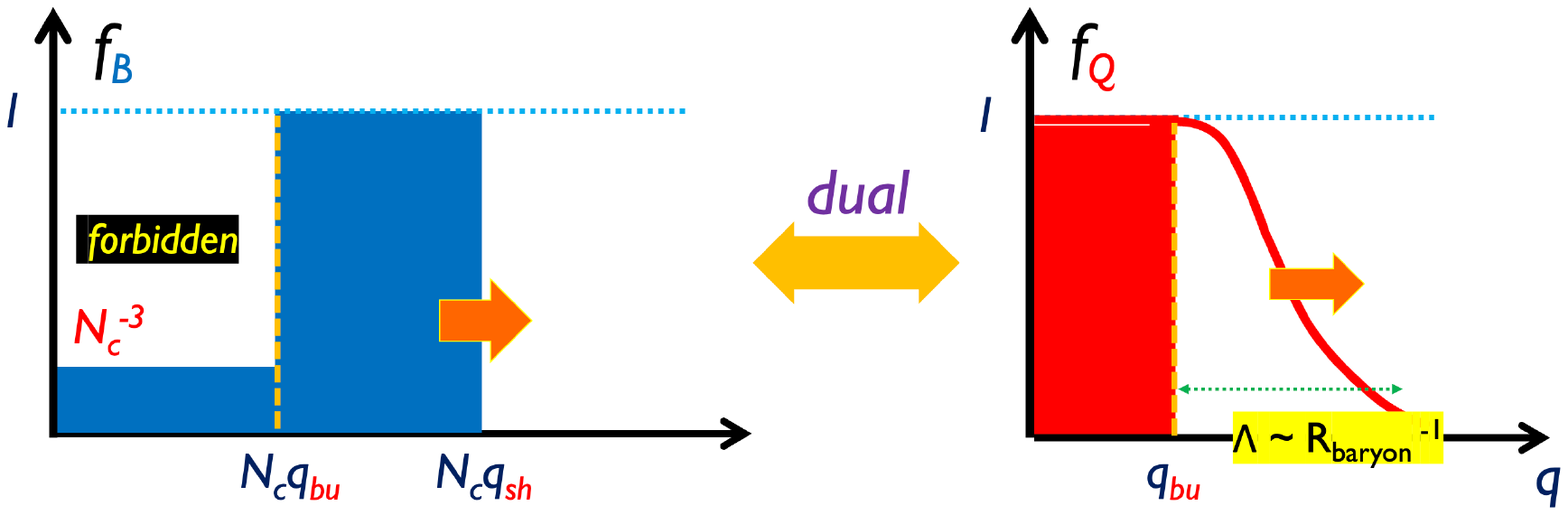}    
        \vspace{-1.4cm}
           \caption{ $f_B$ and $f_Q$ after quark saturation.
    }
    \label{fig:H_to_Q_EOS}
\end{figure}

Now the rest is straightforward; we minimize $\varepsilon$ by optimizing $f_B$ at each momentum.
While the logic is simple, the methodology warrants clarification.
To perform the minimization subject to a fixed baryon density $n_B$, 
we must impose this constraint during the variation of $f_B (k)$.
A practical way to enforce this is to consider paired variations in momentum space such that
%
\begin{eqnarray}
\delta f_B ({\bf k}_1) + \delta f_B ({\bf k}_2) = 0 \,.
\end{eqnarray}
This condition ensures that the total baryon density remains unchanged during the variation. 
The corresponding change in the energy density is then given by
%
\begin{align}
\delta \varepsilon 
&= E_B (k_1) \delta f_B ({\bf k}_1) + E_B (k_2) \delta f_B ({\bf k}_2) 
\notag \\
&= \big[\, E_B (k_1) - E_B (k_2) \, \big] \delta f_B ({\bf k}_1) \,. 
\end{align}
This condition implies that relocating a particle from ${\bf k}_2$ to ${\bf k}_1$ reduces the total energy if $|{\bf k}_2| > |{\bf k}_1|$.
Consequently, the optimal distribution $f_B (k)$ must concentrate particles at low momenta.
We define $k_{\rm sh}$ as the largest momentum below which $f_B$ is nonzero.
Then, we conclude that $f_B (k)$ vanishes for $k > k_{\rm sh}$,
while it is maximized for $k \le k_{\rm sh}$.
The maximum value that $f_B$ can attain is governed by the sum rule constraint.
If no quark momentum states are saturated, then $f_B$ can reach the maximum value of 1, 
recovering the standard ideal gas distribution, $f_B^{\rm ideal} (k) = \Theta (k_{\rm sh}-k)$.
However, when some domain in quark momentum space is saturated, 
the sum rule enforces a bound on $f_B$, limiting it to a maximum of $1/N_c^3$.
This reflects the duality between the saturation of quark states and the suppression of baryonic occupation probabilities.

The final question is how to smoothly patch together the regions with 
different behaviors of the baryon momentum distribution $f_B$,
namely the saturated region $f_B (k) = 1/N_c^3$, the free region $f_B (k) = 1$,
and the unoccupied region $f_B (k) = 0$.
The appropriate form turns out to be
%
\begin{eqnarray}
f_B(k) = \frac{1}{\, N_c^3 \,} \Theta (k_{\rm bu} - k) + \Theta ( k_{\rm sh} - k ) \Theta ( k - k_{\rm bu} ) \,.
\end{eqnarray}
The domain $k \le k_{\rm bu}$ is dual to $f_Q =1$. 
For the domain $k_{\rm bu} \le k \le k_{\rm sh}$,
baryons are free from the quark saturation constraint and hence $f_B$ can reach $1$, the maximum.
When $k > k_{\rm sh}$, $f_B (k)$ drops to zero.

The above $f_B (k)$ is dual to (we define $N_c q_{\rm bu} \equiv k_{\rm bu}$ and $N_c q_{\rm sh} \equiv k_{\rm sh}$) 
\begin{align}
f_Q(q) 
& = \Theta (q_{\rm bu} - q) 
\notag \\
& + f_Q^{f_B=1} (q) \Theta ( q_{\rm sh} - q ) \Theta ( q - q_{\rm bu} ) 
\notag \\
& + f_Q^{f_B=0} (q) \Theta ( q - q_{\rm sh} ) \,.
\end{align}
The solution of $f_Q$ in the region where $f_B =0$ can be expressed as 
a linear combination of two basis functions $g_{\pm}$ that satisfy the homogeneous equation $L [g_\pm] = 0$.
Explicitly the form is $ f_Q^{f_B=0} = c_+ g_+ (q) + c_- g_- (q)$.
In contrast, in the region where $f_B =1$, the solution is given by $ f_Q^{f_B=0} = d_+ g_+ (q) + d_- g_- (q) + N_c^3$.
Here, the large constant term of $O(N_c^3)$ must be canceled 
by the contributions from the homogeneous solutions to ensure physical consistency, 
particularly with the Pauli exclusion principle.
We have four coefficients, but the condition $f_Q (q\rightarrow \infty) \rightarrow 0$ allows us to eliminate one coefficient, $c_+$.
Consequently, we are left with three free parameters, along with the matching scale
 $q_{\rm bu}$ (or equivalently $k_{\rm bu}$), 
 which must be determined by imposing continuity and the sum rule constraint across the different domains of the distribution.

Applying the operator $L$ to the expression of $f_Q$, 
we find that the continuity of $f_Q$ and its first derivate at the matching points $q_{\rm sh}$ and $q_{\rm bu}$ is required.
 If this continuity is not satisfied, the condition $0\le f_B \le 1$ would be violated by the $\delta$-functions and their derivatives in the expression of $f_B$.
 To avoid such unphysical contributions, the coefficients in front of these $\delta$- and $\delta'$-functions must cancel.
 This requirement yields two matching conditions (continuity and differentiability) at each boundary point,
  $q_{\rm bu}$ and $q_{\rm sh}$, resulting in four constraints in total.
These are sufficient to determine all four unknowns—three coefficients in the ansatz for $f_Q$,
and the boundary momentum  $q_{\rm bu}$. 
The explicit construction and solution of these conditions are provided in \cite{Fujimoto:2023mzy}. 

We have just derived the momentum shell solution for $f_B$ 
which is dual to a quark Fermi sea with a diffused Fermi surface in $f_Q$.
In this regime, matter is characterized by a dense quark Fermi sea in the bulk, while baryons define the Fermi surface.
This construction provides a concrete realization of the quarkyonic matter concept originally proposed in \cite{McLerran:2007qj}.
The idea that baryons occupy a momentum shell—surrounding a quark core—was
first conjectured in \cite{McLerran:2018hbz} as a mechanism to achieve rapid stiffening of the equation of state (EOS).
For seminal works in this direction, see, e.g., \cite{Jeong:2019lhv,Duarte:2020xsp,Duarte:2020kvi,Zhao:2020dvu}.

It is often argued that the low momentum part should be dominated by baryons, while quarks should prevail at high momentum,
see, for instance, the discussion in \cite{Koch:2022act}. 
However, our model calculations reveal the opposite trend.
The essential factor is not the momentum scale itself, 
but rather the availability of phase space for forming spatially localized composite particles. 
This viewpoint offers a more physically grounded interpretation of the results obtained in the IdylliQ model.

At high density, the formation of composite particles requires appropriate superpositions of multiple quark quantum states, 
arranged to minimize the total energy.
However, such configurations are feasible only when ample phase space is available to support the required superpositions. 
As quark states become increasingly occupied, this freedom is lost. 
Once the quark phase space is saturated, a baryonic description ceases to be natural.


In contrast, near the Fermi surface, 
ample phase space remains available, allowing for optimized superpositions of quark states.
It is important to note that the saturated quark Fermi sea is color-singlet due to the complete occupation of all color states. 
However, in the unsaturated region, not all superpositions of quark states automatically satisfy the color-singlet constraint.
A natural class of color-singlet configurations near the Fermi surface is provided by baryonic states.
Whether these baryonic excitations further organize into more complex paired states remains an open question for future analysis.

\subsection{Equations of state }

\begin{figure}
    \centering
    \includegraphics[width=0.85\linewidth]{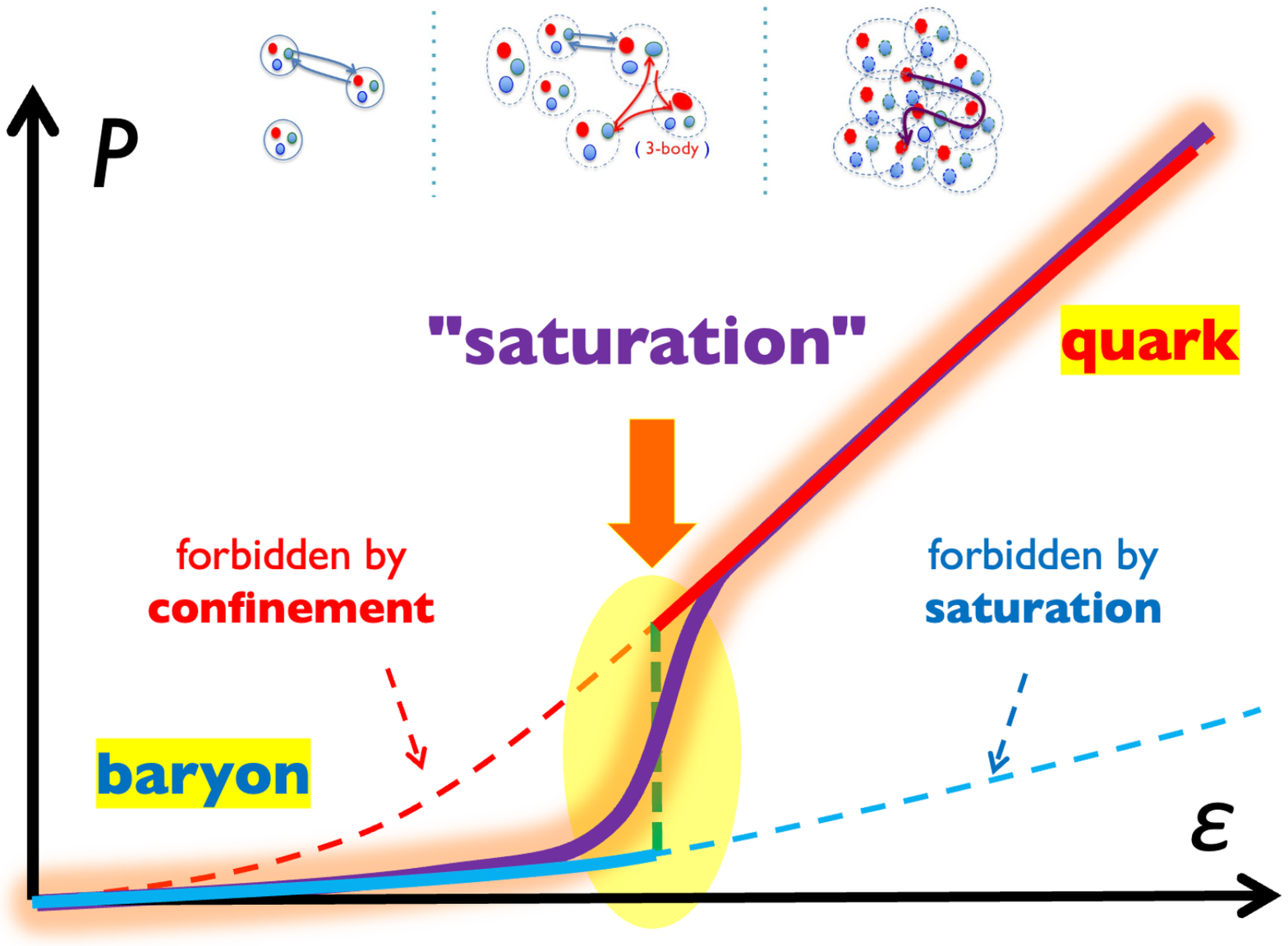}
    \caption{
    The stiffening of the equation of state is associated with the transition from baryonic to quark matter, which is triggered by quark saturation. 
    Baryon-baryon interactions, mediated by quark exchanges, smooth out this transition.
    }
    \label{fig:H_to_Q_EOS}
\end{figure}

EOS in quarkyonic matter rapidly stiffens from the pre- to post-saturation regime.
Before quark saturation sets in, baryonic matter remains largely non-relativistic.
However, once saturation occurs, quarks fill the low-momentum phase space, forcing baryons into higher-momentum states, thereby making them relativistic.
If baryons could occupy momentum states up to the momentum $p_F$ with full probability (i.e., $f_B=1$), 
the relativistic regime $p_F \sim M_B$ would only be reached at a very high density, $n_B \sim M_B^3 \sim 100n_0$.
Taking the quark Pauli blocking constraint into account, however, 
baryons occupy low momentum states only with the probability $\sim 1/N_c^3$. 
As a result, relativistic baryons emerge at much lower densities, 
around $n_B \sim M_B^3/N_c^3 \sim \Lambda^3 \sim 5n_0$.

The explanation of stiff EOS within baryonic descriptions appears highly exotic.
In fact, generating such baryon distributions from conventional nuclear 
models—even with the inclusion of strong many-body repulsions—seems extremely challenging.
As an alternative, a quark-based description offers a more natural path to capture the evolution of stiffness. 
In this approach, we assume the baryon energy can be written as
%
\begin{eqnarray}
E_B(k) = N_c \int_q E_Q ({\bf q}) \varphi ({\bf q}- {\bf k}/N_c) \,.
\end{eqnarray}
where $E_Q ({\bf q})$ represents the single-quark energy and $\varphi$
is the momentum distribution characterizing the internal structure of baryons. 
In the IdylliQ model one can derive the expression of $E_Q$ explicitly by applying the operator $L$,
but we proceed with the present abstract form.

As an example, let us consider the non-relativistic baryon energy.
Expanding in powers of $k/N_c$, we obtain 
\begin{eqnarray}
E_B(k) = N_c \langle E_Q \rangle|_{k=0} + \bigg\langle\frac{\, \partial^2 E_Q \,}{\, \partial q_i \partial q_j} \bigg\rangle \bigg|_{k=0} \frac{\, k_i k_j \,}{2N_c} + \cdots \,,
\end{eqnarray}
where we defined $\langle O \rangle \equiv \int_q O \varphi ({\bf q})$.
The first term corresponds to the baryon mass, 
and the second gives the baryon kinetic energy along with some corrections.
The term linear in $k$ vanishes due to angle averaging.

Using the sum rule and the energy density from Eq.~\eqref{eq:en_fB},
we can express the energy density as
\begin{align}
\varepsilon 
&= N_c \int_{k,q} E_Q (q) \varphi ({\bf q}-{\bf k}/N_c) f_B(k)
\notag \\
&= N_c \int_q E_Q (q) f_Q(q) \,.
\end{align}
Before the quark saturation, $f_Q (q)$ scales as $\sim n_B \varphi (q)$ (see the previous section),
which increases the amplitude but does not alter the typical momentum scale.
Then, $\varepsilon \sim n_B N_c \int_q E_Q \varphi \sim n_B M_B$, as expected.
Here, the energy per baryon scales as $\varepsilon/n_B \sim O(1/N_c)$,
 so the pressure is small, $P \sim O(1/N_c)$.
After saturation,  $f_Q$ behaves as
\begin{eqnarray}
f_Q (q) = \Theta (q_{\rm bu} - q) + f_Q^{\rm sh} (q) \,.
\end{eqnarray}
where $f_Q^{\rm sh} (q) \propto \Theta ( q - q_{\rm bu} ) $.
Unlike the pre-saturation regime,
the evolution of $f_Q$ now proceeds toward higher momentum.
The corresponding energy density is given by
\begin{align}
&\hspace{-0.52cm}
\varepsilon 
= N_c \int_q E_Q (q) \Theta (q_{\rm bu} - q) 
+ N_c \int_q E_Q (q) f_Q^{\rm sh} (q) 
\notag \\
&\hspace{-0.4cm}
=
N_c \int_q E_Q (q) \Theta (q_{\rm bu} - q) 
+ N_c \bar{E}^{\rm sh}_Q \big( n_B -n_{\rm bu} \big)
\,,
\end{align}
where $n_{\rm bu} = \int_k \Theta (q_{\rm bu}-q)$ and $\bar{E}^{\rm sh}_Q \sim E_Q (q_{\rm bu})$ is the average energy for $q \ge q_{\rm bu}$.
Now we examine the energy per baryon, $\varepsilon/n_B$, to study the pressure.
We first consider the case just after the saturation, where the majority of $n_B$ is carried by the momentum shell component.
The energy per baryon is $\varepsilon/n_B \simeq N_c \bar{E}^{\rm sh}_Q$ with which
\begin{eqnarray}
P \simeq N_c n_B^2 \frac{\, \partial \bar{E}^{\rm sh}_Q \,}{\, \partial n_B \,} \,.
\end{eqnarray}
The pressure is $O(N_c)$, since
the derivative of $\bar{E}^{\rm sh}_Q$ is $O(1)$ as $q_{\rm bu}$ increases with $n_B$.
Alternatively, if most of $n_B$ is carried by $n_{\rm bu}$, 
the energy density scales as in standard quark matter with a degenerate Fermi sea.
In this case, $q_{\rm bu} \sim n_B^{1/3}$, and the pressure remains $O(N_c)$.

The above-mentioned arguments suggest that
the quark saturation inevitably drives a rapid stiffening of EOS,
increasing the pressure from $O(1/N_c)$ to $O(N_c)$.
During this transition, the energy density remains $O(N_c)$;
before the saturation it is primarily carried by the baryon mass,
while after the saturation,
it is dominated by the quark kinetic energy.

We note that the quark kinetic energy is always present,
but before the saturation, it does not contribute to the thermodynamic pressure.
This is because confinement prevents quarks from individually contributing to the thermodynamics;
in other words, the mechanical pressure inside baryons cancel each other out.
However, after the saturation, the quark Fermi sea requires a collective orientation of quark momenta,
allowing quarks to contribute to the pressure.

\section{Summary}

The scenario of rapid stiffening presented in this article is based on the 
observation that quarks can produce a stiff EOS.
This is not entirely too surprising,
as the kinetic energy of quarks can be much larger than that of nucleons.
What complicates the situation is confinement:
it prevents quarks from directly contributing to the pressure,
while still allowing them to contribute to the energy density through the baryon masses.
Eventually, quark saturation effects come into play, making the quark matter scaling of the EOS inevitable.

It is surprising to us that quarks already play very important roles
at densities only slightly above the nuclear saturation density.
For reasonable choices of the scale $\Lambda$ in $\varphi$,
the density at which quark saturation occurs is roughly half the density where baryons begin to spatially overlap.
If baryon overlap is assumed to occur around $\sim $5-6$n_0$, then the quark saturation happens at approximately $\sim$ 2-3$n_0$.
Using lattice results for QCD-like theories such as two-color or isospin QCD \citep{Brandt:2022hwy,Abbott:2023coj,Iida:2024irv},
we have examined the overlap density of diquarks or pions based on their radii, reaching similar estimates \citep{Kojo:2021hqh,Chiba:2023ftg,Kojo:2024sca}.
We also note that the mechanism of soft deconfinement \citep{Fukushima:2020cmk}, characterized by the overlap of meson clouds around baryons,
suggests a transition beyond the nuclear regime at about $\sim 2n_0$.
It seems that all of these arguments are related to a breakdown scale around $\sim 2n_0$ (or possibly even lower density) 
as estimated in the chiral effective theory calculations \citep{Drischler:2020fvz}.
There are also attempts to interpret nuclear saturation properties at $n_0$ as a consequence of quark saturation \citep{Koch:2024qnz,McLerran:2024rvk}.

Considerations based on the quark degrees of freedom also impact our baryonic descriptions involving strangeness.
Nucleons, hyperons, and possible excitations such as the $\Delta$ baryons
share quarks and therefore cannot be treated as independent particles.
In neutron star matter, the large number of neutrons saturates the down-quark states first.
With a saturated down-quark Fermi sea, hyperons such as $\Sigma_{0}$, $\Lambda_0$, and some members 
of the decuplet like $\Delta_+$, $\Delta_0$, $\Delta_-$ become energetically disfavored. 
This is because opening phase space for these particles would require 
removing neutrons at low momenta or placing massive baryons in the high-momentum states.
Quark descriptions mitigate the softening of EOS associated with the appearance of strangeness \citep{Fujimoto:2024doc}.

The current version of the model is still primitive, and many issues remain to be clarified.
To complete the description of the crossover, it is necessary to discuss how baryon-baryon interactions are related to the underlying quark dynamics.
Quark exchanges appear to play the most significant role in baryon interactions; 
at long distances, these manifest as meson exchanges, while at short distances they typically produce a hard-core repulsion. 
Another important aspect to consider is the structural changes of hadrons triggered by quark exchanges.
Further studies are clearly needed to better understand the crossover regime.


\section*{Acknowledgments}

I thank G. Baym and T. Hatsuda with whom the arguments in the first half of this manuscript have been developed.
The content related to quarkyonic matter is based on the collaboration with Y. Fujimoto and L. McLerran.
The estimate for the density of the quark saturation is largely based on the collaboration with
R. Chiba and D. Suenaga.
This work is supported by JSPS KAKENHI Grant No. 23K03377 and No. 18H05407 and by the Graduate Program on Physics for the Universe (GPPU) at Tohoku University.

\bibliographystyle{cas-model2-names}
\bibliography{ref}

\end{document}